\documentclass[useAMS,usenatbib]{mn2e}
\usepackage{times}
\usepackage{epsfig}

\title[A nearby luminous AGN sample from HST]{A nearby luminous AGN sample optically selected from Hubble Space Telescope (HST)}
\author[S.-L. Li]{Shuang-Liang Li \thanks{E-mails:lisl@shao.ac.cn}\\
Key Laboratory for Research in Galaxies and Cosmology, Shanghai Astronomical Observatory, Chinese Academy of Sciences,\\
80 Nandan Road, 200030 Shanghai, China}

\begin{document}

\pagerange{\pageref{firstpage}--\pageref{lastpage}} \pubyear{2019}

\maketitle

\label{firstpage}

\begin{abstract}
In this work, a nearby luminous AGN sample is selected from HST, where only sources with both X-ray emission observed by \textit{Chandra/XMM-Newton} and radio flux detected by VLA/VLBA/VLBI/MERLIN are adopted to keep high precision. We get a sample of 30 luminous AGNs finally, which consist of 11 RLAGN and 19 RQAGN. It is found that the relationship between $R_{\rm {UV}}$ and $\alpha_{\rm {ox}}$, which was firstly reported by \citet{l2017} in LLAGN, and other relationships are all absent in RLAGN, probably due to the complex physical process therein. Our results indicate that the X-ray emission from jet should play an important role in RLAGN and further support the transition of accretion mode between LLAGN and RLAGN. On the other hand, the traditional relationships in RQAGN, such as $\alpha_{\rm {ox}}$ and $\lambda$, $\Gamma$ and $\lambda$, are found to be well consistent with previous works.
\end{abstract}

\begin{keywords}
accretion, accretion discs $-$ black hole physics $-$ galaxies: active $-$ quasars: general
\end{keywords}

\section{INTRODUCTION}

It is generally accepted that active galactic nuclei (AGNs) are powered by the release of gravitational energy as gas falls into the central black holes. With the increase of AGN luminosity, the accretion mode is believed to transfer from a geometrically thick, optically thin and radiatively inefficient accretion flow (RIAF) in low-luminosity AGNs (LLAGN) \citep{n1994,y2014} to an optically thick and radiatively efficient thin disc in luminous AGNs \citep{s1973}. Lots of observations have been achieved to support this transition. For example, the spectral energy distribution (SED) in luminous AGNs exhibits a so-called big blue bump in optical/UV bands, which is absent in LLAGN \citep*[e.g.,][]{e1994,h2009,s2011}. The big blue bump has been regarded as originating from the black body emission of a thin disc. Secondly, a strong positive correlation between the optical-to-X-ray spectral index $\alpha_{\rm ox}$ and Eddington ratio $\lambda$ has been extensively explored and validated in luminous AGNs by numerous works \citep[e.g.,][]{s2006,l2010,j2012,b2015}, while a converse correlation was discovered in LLAGN \citep[e.g.,][]{x2011,l2017}. Furthermore, the correlation between the hard X-ray photon index $\Gamma$ and $\lambda$ is also found to be different for LLAGN and luminous AGNs \citep[e.g.,][]{g2009,b2015}.

The dominant parameter controlling the transition between LLAGN and luminous AGNs should be mass accretion rate in accretion disc. However, the correlations among $\alpha_{\rm ox}$, $\Gamma$ and $\lambda$ in LLAGN all display large scatters, indicating that at least an additional parameter is necessary when exploring the physics of LLAGN. \citet{l2017} suggested that the viscosity parameter $\alpha$ in RIAF can play certain role for the reason that $\alpha$ can significantly affect the structure of RIAF \citep{n1994,m1997}. Moreover, by compiling a sample of 32 LLAGN (including 18 low-ionization nuclear emission-line region galaxies and 14 low luminosity Seyfert galaxies), they found a new strong anti-correlation between the radio loudness $R_{\rm UV}$ and $\alpha_{\rm ox}$. Except for this new relationship, two further relationships reported in the literature ($R_{\rm UV}$-$\lambda$ and $\alpha_{\rm ox}$-$\lambda$ correlations) can be well explained by a truncated accretion disc-jet model (see \citealt{l2017} for details).

In this work, we intend to investigate the correlations among various physical properties, such as $R_{\rm UV}$, $\alpha_{\rm ox}$, $\Gamma$ and $\lambda$ in luminous AGNs, especially the correlation between $R_{\rm UV}$ and $\alpha_{\rm ox}$ to further validate the transition of accretion mode. In order to get the high precision, for the first time, we construct a luminous AGN sample, where the optical emission is observed by HST, the X-ray flux is from \textit{Chandra/XMM-Newton}, and the radio band is gotten by VLA/VLBA/VLBI/MERLIN.

\section{THE SAMPLE}\label{sample}

We search the literature for nearby luminous AGNs where the optical/UV emission is observed by HST. The sources in our sample are all picked up from the following three papers: 1), the low redshift ($z < 0.5$) quasar sample collected from HST archives \citep{p2003}, which contains 34 radio-loud AGNs (RLAGN); 2), the 70 low-redshift ($0.06<z<0.46$) luminous AGNs observed with HST, which includes both radio-quiet and radio-loud sources \citep{h2008b}; 3) the 28 low-redshift ($z < 0.3$) Palomar-Green quasars from a deep HST survey \citep{v2009}.

The radio flux in both RLAGN and LLAGN is believed to come from the synchrotron emission of jet. However, the origin of radio emission in radio-quiet AGNs (RQAGN) is still unclear, which may arise from the corona above accretion disc \citep[e.g.,][]{l2008}, or the star formation \citep[e.g.,][]{z2016}, or the AGN outflows \citep[e.g.,][]{j2010,l2019}, or even a weak jet \citep[e.g.,][]{w1995}. Therefore, in order to compare the correlation of $R_{\rm {UV}}$ and $\alpha_{\rm {ox}}$ between luminous AGNs and LLAGN, we need to separate our sample into two parts, i.e., RLAGN and RQAGN. Generally, AGNs are divided to radio-loud or radio-quiet according to their ratio of radio flux and optical flux, which is defined as $R_{\rm {UV}}=f_{\rm {5GHz}}/f_{\rm {4400{\AA}}}$ \citep{k1989}.

\subsection{RLAGN sample}

\begin{table*}
\normalsize
\flushleft
\caption{The RLAGN sample.}
\begin{minipage}{\textwidth}
\begin{center}
\begin{tabular}{llcccccccc}
\hline
{Name} & {$z$} & {log$M_{\rm
BH}/M_{\odot}$} &  {log$L_{\rm R}$} &
{log$L_{\rm UV}$} &  {log$L_{\rm X}$} &
{log$\lambda$} &  {log$R_{\rm UV}$} &  {$\Gamma$}  &  {$\alpha_{\rm ox}$} \\
{(1)} &  {(2)} &  {(3)} &  {(4)} &
 {(5)} &  {(6)} &  {(7)} &  {(8)}  &  {(9)} & {(10)} \\
\hline

0133+207 &  0.425 &  9.0$^{(1)}$  &  32.43$^\ast$  &    30.39  &  26.70 & -0.75  &    1.96 &   1.28$^{(a)}$ &    1.42 \\
3C 48    &  0.367 &  8.7$^{(1)}$  &  34.18  &    30.86  &  27.17 & 0.021  &    3.24 &   2.12$^{(a)}$ &    1.42 \\
1100+772 &  0.315 &  9.5$^{(1)}$  &  32.15$^\ast$  &    30.92  &  27.02 & -0.72  &    1.16 &   1.77$^{(b)}$ &    1.50 \\
3C 273   &  0.158 &  8.8$^{(2)}$  &  34.13  &    31.39  &  27.75 & 0.45   &    2.67 &   1.86$^{(c)}$ &    1.40 \\
1302-102 &  0.286 &  9.04$^{(3)}$ &  32.89  &    31.02  &  26.79 & -0.16  &    1.79 &   1.66$^{(b)}$ &    1.62 \\
1309+355 &  0.184 &  8.1$^{(1)}$  &  31.40$^\ast$  &    30.31  &  25.88 & 0.07   &    1.01 &   1.94$^{(b)}$ &    1.70 \\
1512+370 &  0.37  &  9.2$^{(1)}$  &  32.27  &    30.67  &  27.09 & -0.67  &    1.53 &   2.15$^{(b)}$ &    1.38 \\
3C323.1  &  0.264 &  8.8$^{(1)}$  &  31.74  &    30.36  &  26.09 & -0.58  &    1.31 &   1.60$^{(d)}$ &    1.64 \\
2135-147 &  0.2   &  9.4$^{(4)}$  &  31.69$^\ast$  &    30.19  &  27.12 & -1.35  &    1.43 &   2.50$^{(e)}$ &    1.18 \\
OX 169   &  0.213 &  8.7$^{(1)}$  &  32.81  &    30.35  &  26.23 & -0.49  &    2.34 &   1.71$^{(f)}$ &    1.58 \\
0903+169 &  0.412 &  8.8$^{(1)}$  &  31.95  &    30.33  &  27.34 & -0.61  &    1.50 &   1.80$^{(g)}$ &    1.15 \\
\hline
\end{tabular}
\end{center}
\end{minipage}
Notes: Col. (1): Source name. Col. (2): Redshift. Col. (3): Black hole mass. Col. (4): radio spectral luminosity at 5 GHz. Col. (5): UV spectral luminosity at 2500 $\rm \AA$. Col. (6): X-ray spectral luminosity at 2 keV. Col. (7): Eddington ratio, $\lambda=L_{\rm {bol}}/L_{\rm {Edd}}$. Col. (8): Radio loudness $R_{\rm UV}\equiv L_{\rm R}/L_{\rm UV}$. Col. (9): Photon index $\Gamma$ at 2-10 kev ($f_\nu \sim \nu^{1-\Gamma}$). Col. (10): optical-to-X-ray spectral index $\alpha_{\rm ox}\equiv 0.384 \log\left[L_{\rm UV}/L_{\rm {X}}\right]$. \\
$^\ast$: For sources labelled with $^\ast$, their radio flux at 5 GHz are derived based on observations at neighbouring frequencies. \\
$^{(1)-(4)}$: References for black hole mass. (1): \citet{d2008}; (2): \citet{s2007b}; (3): \citet{i2007}; (4): \citet{h2008b}.
\\
$^{(a)-(g)}$: References for photon index. (a): \citet{s2005}; (b): \citet{i2007}; (c): \citet{p2005}; (d): \citet{m2010}; (e): \citet{f2005}; (f): \citet{g2010}; (g): \citet{h2006}.
\end{table*}

Our radio-loud AGN sample is selected from \citet{p2003}, which includes 34 radio-loud quasars with redshift $z<0.5$ and absolute blue magnitude $M_{\rm B}<-23$. The 34 radio-loud quasars are composed of two parts, i.e., 11 new objects through searching HST data with the Wide Field Planetary Camera 2 (WFPC2), and 23 old objects gathered from other literatures (see \citealt{p2003} for details). We search the nuclear radio luminosity at 5 GHz of their sample from NED\footnote{website: http://ned.ipac.caltech.edu/}, and find that only 22 objects had been observed by high resolution VLA/VLBA/VLBI/MERLIN. Furthermore, 10 sources are reduced because their X-ray fluxes aren't observed by \textit{Chandra/XMM-Newton}. Getting rid of one another object 0903+169 (its black hole mass is unavailable), we obtain 11 objects finally (see table 1). All the RLAGN in table 1 have Eddington ratio larger than 0.01, indicating their central engines should be an optically thick thin disc. According to the traditional classification of radio galaxies based on jet morphology, most of our sources are FRII radio galaxies \citep{f1974}, which usually possess higher jet power, edge-brightened radio lobes and hot spots. Therefore, we have to constrain on the nuclear region in order to get the nuclear emissions. We notice that our sample is quite small, however, we still suggest that it can relatively represent the population of RLAGN for the reason that all the parameters ($\lambda$, $\Gamma$, $\alpha_{\rm {ox}}$, etc) cover almost the same range as those of RQAGN (see Fig.1 and Fig.2).       

Col. (1), (2) in Table 1 are directly taken from \citet{p2003}. Though the black hole mass in Col. (3) is taken from five different references in this work (see table 1 and 2 for details), it is all consistently determined through the virial relation between black hole mass and $H_\beta$ line width. We collect the nuclear radio luminosity at 5 GHz from NED as shown in col. (4). For sources without radio observations at 5 GHz (labelled with superscript $^\ast$), their radio luminosity at 5 GHz are derived from observations at neighbouring frequencies, with a radio spectral index $\alpha_{\rm r}\approx -0.5$ ($f_\nu \sim \nu^{\alpha_{\rm r}}$, e.g., \citealt{s2011}). In col. (5), we present the optical-UV spectral luminosity at 2500${\rm \AA}$ by extending the optical luminosity at R band in \citet{p2003} with a spectral index $\alpha_{\rm o}\approx -0.5$ ($f_\nu \sim \nu^{\alpha_{\rm o}}$, e.g., \citealt{h2008,s2011}). The Eddington ratio $\lambda$ in col. (7) is defined as $\lambda=L_{\rm {bol}}/L_{\rm {Edd}}$, where the bolometric luminosity $L_{\rm {bol}}$ is calculated with $L_{\rm {bol}}=8.1 \lambda L_{\rm \lambda}$ ($\lambda=5100 {\rm \AA}$, \citealt{r2012}), and $L_{\rm {Edd}}=1.3 \times 10^{38}(M_{\rm {BH}}/M_\odot)$ erg s$^{-1}$ is the Eddington luminosity. We notice that \citet{h2008b} also gave a radio-loud AGN sample observed by HST. However, their sample is totally overlapped with that in \citet{p2003} when both the X-ray and radio luminosity are available.

\subsection{RQAGN sample}

\begin{table*}
\normalsize
\flushleft
\caption{The RQAGN sample.}
\begin{minipage}{\textwidth}
\begin{center}
\begin{tabular}{llccccccc}
\hline
{Name} & {$z$} & {log$M_{\rm {BH}}/M_{\odot}$} &
{log$L_{\rm UV}$} &  {log$L_{\rm X}$} &
{log$\lambda$} &    {$\Gamma$}  &  {$\alpha_{\rm ox}$} &  {$\kappa_{\rm {bol}}$}\\
{(1)} &  {(2)} &  {(3)} &  {(4)} &
 {(5)} &  {(6)} &  {(7)} &  {(8)}  &  {(9)} \\
\hline

PG0050+124$^{(II)}$ &  0.061 &  7.26$^{(1)}$  &  29.46  &    26.11  &  -0.11 & 2.28$^{(a)}$  &    1.28 &   12.62 \\
PG0052+251$^{(I)}$ &  0.155 &  8.72$^{(1)}$  &  30.05  &    26.89  &  -1.05 & 1.83$^{(c)}$  &    1.25 &   6.91 \\
PG0157+001$^{(II)}$ &  0.163 &  8.03$^{(1)}$  &  30.22  &    26.12  &  -0.12 & 2.10$^{(a)}$  &    1.59 &   67.01 \\
PG0844+349$^{(II)}$ &  0.064 &  7.8$^{(2)}$   &  29.40  &    26.02  &  -0.71 & 2.11$^{(a)}$  &    1.32 &   13.75 \\
PG0953+414$^{(I)}$ &  0.234 &  8.58$^{(1)}$  &  30.62  &    26.97  &  -0.34 & 2.01$^{(a)}$  &    1.41 &   20.13 \\
PG1116+215$^{(II)}$ &  0.176 &  8.64$^{(2)}$  &  30.32  &    26.75  &  -0.63 & 2.14$^{(a)}$  &    1.53 &   19.49 \\
PG1202+281$^{(I)}$ &  0.165 &  8.54$^{(2)}$  &  29.74  &    26.74  &  -1.18 & 1.69$^{(a)}$  &    1.21 &   5.01 \\
PG1216+069$^{(I)}$ &  0.331 &  9.17$^{(1)}$  &  30.83  &    26.97  &  -0.72 & 1.73$^{(a)}$  &    1.54 &   35.15 \\
PG1307+085$^{(II)}$ &  0.155 &  8.73$^{(1)}$  &  29.89  &    26.40  &  -1.15 & 1.46$^{(a)}$  &    1.46 &   18.34 \\
PG1402+261$^{(I)}$ &  0.164 &  7.94$^{(2)}$  &  30.18  &    26.43  &  -0.14 & 2.06$^{(a)}$  &    1.44 &   25.61 \\
PG1411+442$^{(II)}$ &  0.09  &  7.98$^{(2)}$  &  29.76  &    25.70  &  -0.53 & 1.90$^{(c)}$  &    1.61 &   64.03 \\
PG1416-129$^{(I)}$ &  0.129 &  8.5$^{(1)}$   &  29.88  &    26.43  &  -1.00 & 1.54$^{(c)}$  &    1.41 &   13.84 \\
PG1426+015$^{(I)}$ &  0.086 &  8.75$^{(1)}$  &  30.08  &    26.44  &  -1.05 & 1.99$^{(c)}$  &    1.40 &   20.41 \\
PG1440+356$^{(II)}$ &  0.079 &  7.3$^{(1)}$   &  29.44  &    25.90  &  -0.17 & 2.03$^{(a)}$  &    1.39 &   19.44 \\
PG1444+407$^{(I)}$ &  0.267 &  8.36$^{(2)}$  &  30.58  &    26.41  &  -0.16 & 2.12$^{(a)}$  &    1.59 &   65.43 \\
PG1613+658$^{(II)}$ &  0.129 &  8.99$^{(1)}$  &  29.99  &    26.54  &  -1.31 & 1.70$^{(a)}$  &    1.41 &   16.03 \\
PG1626+554$^{(II)}$ &  0.133 &  8.43$^{(2)}$  &  29.84  &    26.44  &  -0.90 & 1.95$^{(a)}$  &    1.35 &   14.02 \\
PG2130+099$^{(II)}$ &  0.063 &  7.68$^{(1)}$  &  29.41  &    25.80  &  -0.58 & 1.65$^{(b)}$  &    1.47 &   23.74 \\
PG2214+139$^{(II)}$ &  0.066 &  8.38$^{(2)}$  &  29.45  &    26.09  &  -1.24 & 1.91$^{(c)}$  &    1.34 &   13.19 \\
\hline
\end{tabular}
\end{center}
\end{minipage}
Notes: The meaning of col. (1)-(9) are the same as those in table 1, col. (10): bolometric correction
$\kappa_{\rm {bol}} =  L_{\rm {bol}}/L_{\rm {[2-10] keV}}$.\\
$^{(I),(II)}$: Here the superscripts (I) and (II) represent sources that are taken from \citet{h2008b} and \citet{v2009}, respectively.\\
$^{(1),(2)}$: References for black hole mass. (1): \citet{v2002}; (2): \citet{i2007}.
\\
$^{(a)-(c)}$: References for photon index. (a): \citet{p2005}; (b): \citet{i2007}; (c): \citet{z2010}.
\end{table*}

Our RQAGN sample is gathered from \citet{h2008b} and \citet{v2009}. The sample in \citet{h2008b} is composed by 70 low redshift luminous AGNs observed by HST WFPC2, with absolute magnitudes brighter than $M_{\rm V}< -23$. Reduced the radio-loud sources, we obtain 8 sources where X-ray luminosity, photon index, and black hole mass are all available, as shown in table 2. Similarly, 11 radio-quiet sources are collected from \citet{v2009}. Our final sample contains 19 RQAGN (see table 2), where col. (1)-(9) are similarly acquired as table 1.

\section{RESULTS}\label{results}

\begin{figure*}
\includegraphics[width=16cm]{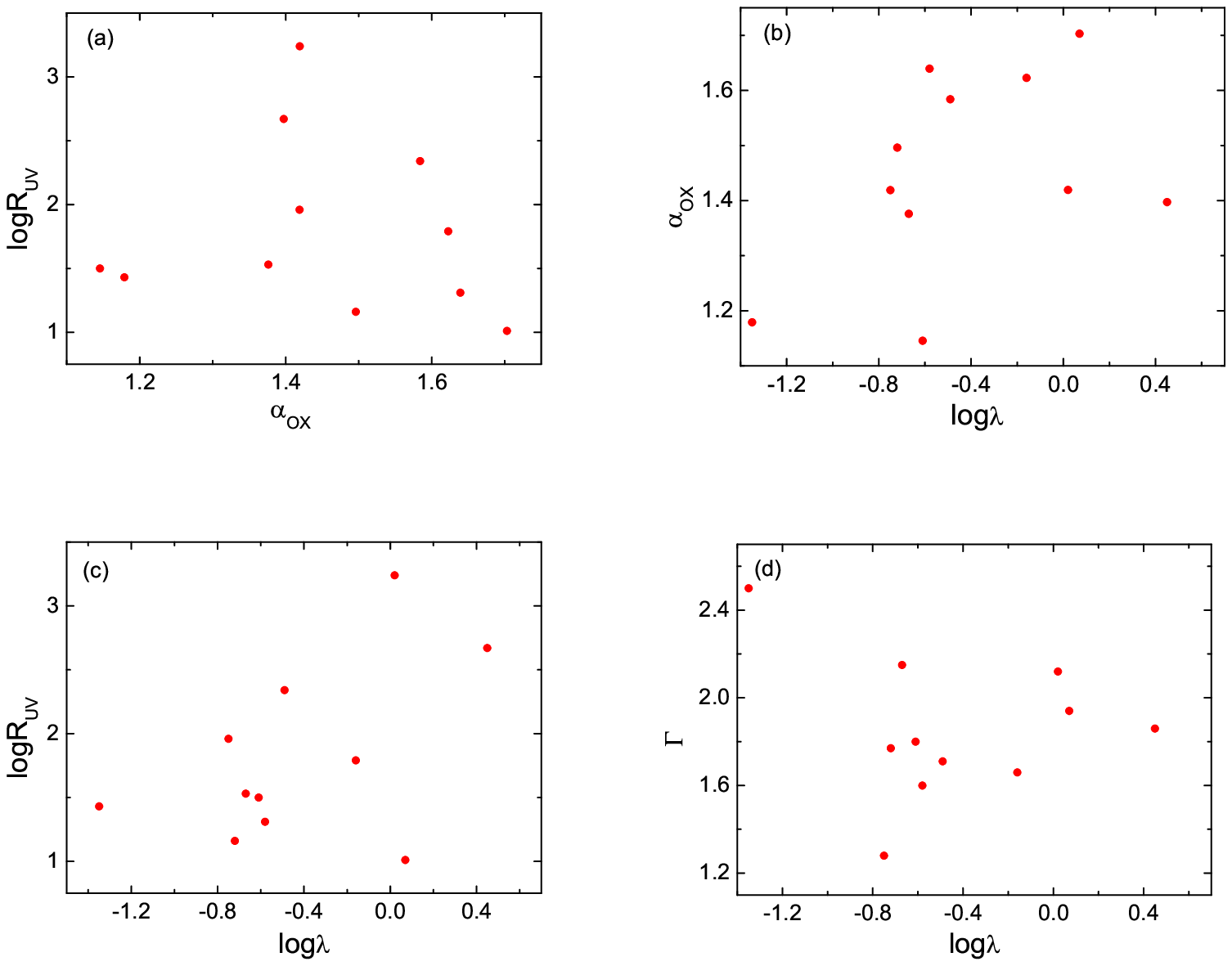}
\caption{The various relationships in RLAGN. \label{loud}}
\end{figure*}

\begin{figure*}
\includegraphics[width=16cm]{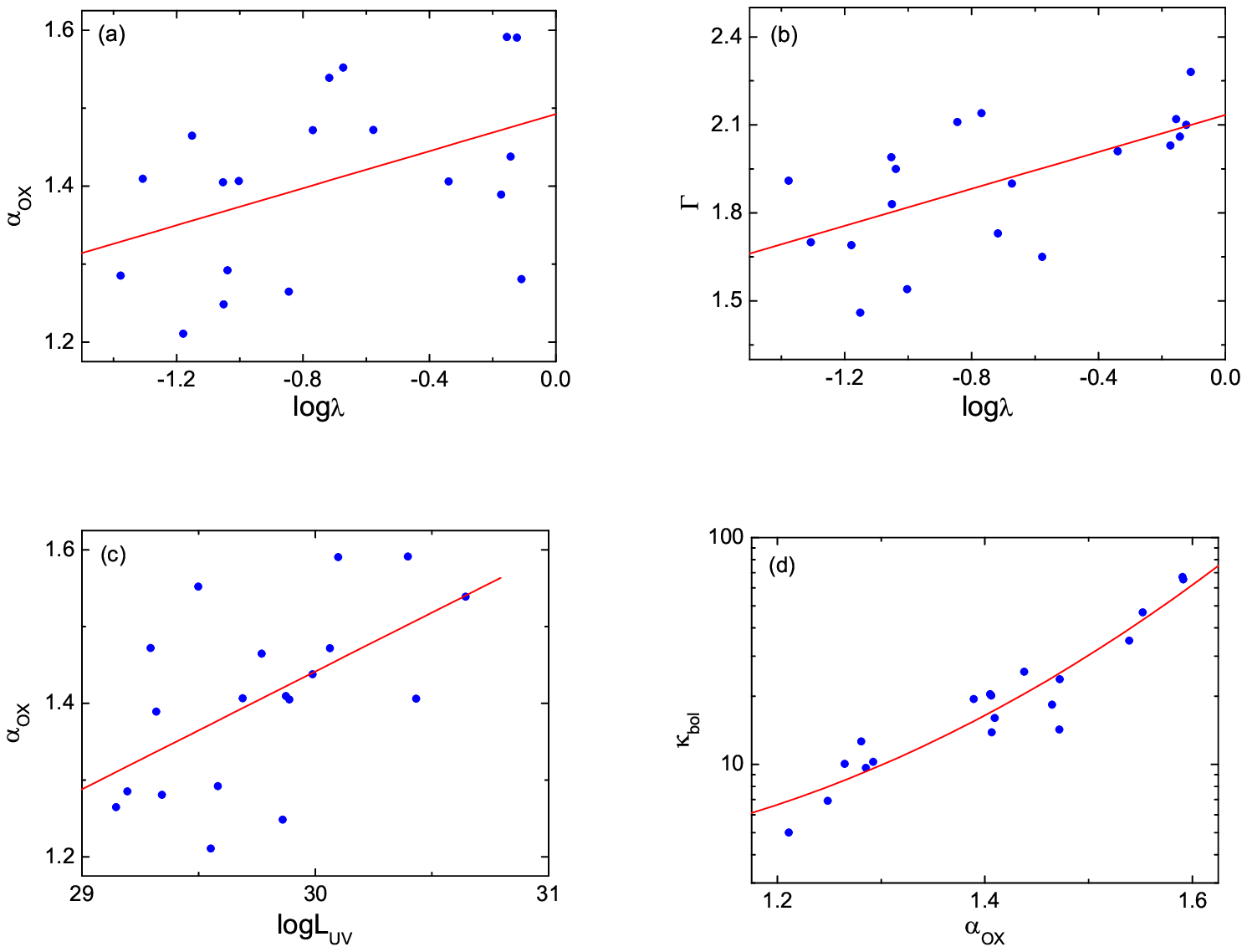}
\caption{The various relationships in RQAGN. \label{quiet}}
\end{figure*}

\subsection{Results for RLAGN}\label{RL results}

\citet{l2017} found a new strong correlation between $R_{\rm {UV}}$ and $\alpha_{\rm {ox}}$ in LLAGN, which can be well understood with the truncated accretion disc-jet model \citep{y2014}. In order to compare with that in LLAGN, we firstly investigate the correlation of $R_{\rm {UV}}$ and $\alpha_{\rm {ox}}$ in RLAGN as shown in Fig. \ref{loud}(a).
However, no significant correlation is found between $R_{\rm {UV}}$ and $\alpha_{\rm {ox}}$, which can provide further verification for the accretion mode transition between RLAGN and LLAGN. Fig. \ref{loud}(b), (c), (d) investigate the correlations between $\alpha_{\rm {ox}}$ and $\lambda$, $R_{\rm {UV}}$ and $\lambda$, $\Gamma$ and $\lambda$, respectively. Similarly, we don't detect any significant correlations in these figures, in contrast with the negative correlations between $\alpha_{\rm {ox}}$ and $\lambda$, and between $\Gamma$ and $\lambda$ in LLAGN \citep[e.g.,][]{g2009,x2011}.

The reason for these discrepancies are still poorly understood. Lots of factors can change the results of observations, such as: firstly, the radio emission in LLAGN may originate from a conical jet, resulting a radio flux $L_{\rm R}\sim \dot{m}^{1.4}$ (e.g., \citealt{m2003}), while the radio flux may be proportional to the accretion luminosity in luminous AGNs (e.g., \citealt{g2014}). The variation in radio emission can affect the dependance of $R_{\rm {UV}}$ on mass accretion rate $\dot{m}$.

For the second, the X-ray flux in LLAGN should come from a radiatively inefficient accretion flow \citep{y2014,l2018}. However, the origin of X-ray in luminous AGNs is still debatable. Generally, the x-ray emission from jet is believed to be important in RLAGN (e.g., \citealt{w1987,w2013}). However a recent work by \citet{g2018} suggested the X-ray emission in both radio-loud and radio-quiet AGNs are emerged from a same region and by a same mechanism, because their X-ray spectral slopes are very similar. Our results suggest that the X-ray flux contributed by jet should be very important, for the reason that there are not significant not only between $R_{\rm {UV}}$ and $\alpha_{\rm {ox}}$, but also between $\alpha_{\rm {ox}}$ and $\lambda$, $\Gamma$ and $\lambda$. The latter two relationships are found to be very strong in RQAGN (see Fig. \ref{quiet}).

Lastly, large-scale magnetic field is necessary in order to accelerate a collimated jet \citep{b1977,b1982}. However, the formation mechanism of magnetic field is still debatable. One possible way is that an initial weak magnetic field can be effectively dragged inwards from outer boundary of accretion disc (e.g., \citealt{l1994,t2011,c2013, b2016,b2018}). Besides, the presence of magnetic field can not only produce jet, but also influence the emission of optical and X-ray \citep[e.g.,][]{q2013,l2014,b2019a,b2019b}. All these factors will change the physical process in RLAGN.

\subsection{Results for RQAGN}

In Figure \ref{quiet}, we explore the various relationships among the physical parameters in RQAGN. A relatively strong correlation between $\alpha_{\rm {ox}}$ and $\lambda$ is found (see Fig. \ref{quiet}a), which reads
\begin{equation}
\alpha_{\rm {ox}}=1.49\pm0.05+(0.12\pm0.06)\log\lambda,
\end{equation}
where the confidence level based on a Pearson test is $\sim 94.3\%$. This is well consistent with previous results (e.g., \citealt{g2010b,l2010}, where the slopes are 0.11 and 0.13, respectively).

In Fig. \ref{quiet}(b), (c), (d), the relationships between $\Gamma$ and $\lambda$, $\alpha_{\rm {ox}}$ and $L_{\rm {UV}}$, $\kappa_{\rm {bol}}$ and $\alpha_{\rm {ox}}$ are given by
\begin{equation}
\Gamma=2.13\pm0.08+(0.31\pm0.10)\log\lambda,
\end{equation}
\begin{equation}
\alpha_{\rm {ox}}=-3.16\pm1.61+(0.15\pm0.05)\log L_{\rm {UV}},
\end{equation}
\begin{equation}
\log\kappa_{\rm {bol}}=2.75\pm3.75-(3.91\pm5.36)\alpha_{\rm {ox}}+(2.26\pm1.91)\alpha_{\rm {ox}}^2, \label{kbol}
\end{equation}
respectively, with confidence level $\sim 99.5\%, 98.9\%$, and $99.9\%$, respectively. These results are also qualitatively consistent with other works (e.g., \citealt{s2006,r2009,g2010b,l2010,b2013b}). We adopt a quadratic fit for the relationship between $\kappa_{\rm {bol}}$ and $\alpha_{\rm {ox}}$ in Equation (\ref{kbol}), simply following \citet{l2010}. The authors found the import of a quadratic term can significantly improves the fit quality. We also give a linear fit for them in this work, which reads
\begin{equation}
\log\kappa_{\rm {bol}}=-2.18\pm0.3+(2.44\pm0.21)\alpha_{\rm {ox}},
\end{equation}
with confidence level $> 99.9\%$.

In contrast to RLAGN, the physical process in RQAGN is somewhat clear. Regardless of the radio emission, the central engine of RQAGN can be simply described as a disc-corona model, where the optical-UV emission originates from a geometrically thin and optically thick accretion disc, while the hard X-ray is produced by the hot corona above disc through inverse Compton scattering of seed photons from disc \citep{j2012,q2013,l2017b}. With the variation of mass accretion rate, the correlations in Fig. \ref{quiet} can be roughly reproduced. The theoretical study on RQAGN strongly dependant on the the properties of hot corona, which is still quite unclear so far. Usually, corona is very compact and small, of the order tens or less gravitational radii ($r_{\rm g}=GM/C^2$, e.g., \citealt{u2014}). But future work is necessary to further explore its geometry, composition, etc.

\section{SUMMARY}\label{summary}

In this work, we compile a sample of nearby luminous AGNs optically selected from HST. In order to get high precision, only sources with both X-ray emission observed by \textit{Chandra/XMM-Newton} and radio emission observed by VLA/VLBI/VLBA/MERLIN are adopted, resulting in a sample of 30 luminous AGNs, including 11 RLAGN and 19 RQAGN. The strong relationship between $R_{\rm {UV}}$ and $\alpha_{\rm {ox}}$ found in LLAGN \citep{l2017} isn't present in our RLAGN sample (see Fig. \ref{loud}), probably due to the complex physical process therein. Besides, we don't find any significant relationship in RLAGN, indicating that the contribution by jet should be very important in X-ray emission of RLAGN (see section \ref{RL results}). Our results can further verify the accretion mode transition between RLAGN and LLAGN.

On the contrary, lots of relationships are obtained in the RQAGN sample, i.e., the relationships between $\alpha_{\rm {ox}}$ and $\lambda$, $\Gamma$ and $\lambda$, $\alpha_{\rm {ox}}$ and $L_{\rm {UV}}$, $\kappa_{\rm {bol}}$ and $\alpha_{\rm {ox}}$ (see Fig. \ref{quiet}). All these relationships are qualitatively consistent with previous results and can be understood under the disc-corona model.

\section*{Acknowledgements}

We thank the reviewer for helpful comments. SLL thank M. Gu for helpful comments and discussion. This work is supported by the NSFC (grants 11773056) and the Key Research Program of Frontier Sciences of CAS (no. QYZDJ-SSW-SYS023). This work has made extensive use of the NASA/IPAC Extragalactic Database (NED), which is operated by the Jet Propulsion Laboratory, California Institute of Technology, under contract with the National Aeronautics and Space Administration (NASA).

\label{lastpage}

\end{document}